\documentclass[12pt,preprint]{aastex}
\usepackage{natbib}
\usepackage{graphicx}
\shorttitle{Atmospheric absorption of dark photon dark matter}
\begin{document}
\title{Atmospheric absorption of dark matter}
\author{Man Ho Chan}
\affil{Department of Science and Environmental Studies, The Education University of Hong Kong, Hong Kong, China}
\email{chanmh@eduhk.hk}

\begin{abstract}
Typically, the interaction between dark matter and ordinary matter is assumed to be very small. Nevertheless, in this article, I show that the effective resonant absorption of dark photon dark matter in the atmosphere is definitely possible. This might also be associated with the alleged temperature anomalies observed in our upper stratosphere. By allowing a small amount of additional energy deposition to our upper stratosphere, a narrow dark matter mass range $m_A \sim 0.0001-0.001$ eV and the corresponding range of the mixing parameter $\varepsilon$ are constrained for the first time. This proposal might overturn our usual assumption of extremely weak interaction between dark matter and ordinary matter and revive the hope of detecting dark matter directly. Some important implications of this proposal such as the heating of planets and supermassive dark stars would also be discussed. 
\end{abstract}


\section{Introduction}
The nature of dark matter is a long-lasting mystery in cosmology. For the most popular candidate Weakly Interacting Massive Particles (WIMPs), many direct-detection experiments (e.g. PICO-60, XENON1T, LUX-ZEPLIN, DEAP-3600) are trying to detect if there is any interaction between WIMP dark matter particles and ordinary matter \citep{Amole,Aprile,Aprile3,Aprile2,Akerib,Aalbers,Adhikari}. However, the null detection of dark matter signal makes the problem much more severe. A large parameter space of interaction cross section against dark matter mass has been ruled out, which gives significant tension with the predicted properties of WIMPs from particle physics \citep{Aprile2,Aalbers,Arcadi}. For indirect-detection of WIMP dark matter, no compelling signal has been received either \citep{Ackermann,Albert,Aguilar,Chan,Beck}. Besides, for another popular candidate axion dark matter, some cavity experiments using haloscopes (e.g. ADMX, CAPP, RADES) are also trying to detect if there is any axion-photon conversion signal under a strong magnetic field \citep{Kwon,Adair,ADMX,Yang,Ahyoune}. Nevertheless, no such photon signal has been recorded for a large range of axion mass. Although the above results do not rule out the possibility of WIMPs or axions being the major component of cosmological dark matter, they have increased the crisis of incorporating cold dark matter into the standard cosmological model.

Recently, \citet{Zioutas} reported that the stratospheric temperature shows a strong peak around December-January in each year between 1986-2018, which is not correlated with solar activity \citep{Zioutas,Zhitnitsky}. Further analysis with the temperature fluctuation in the stratosphere demonstrates a possible correlation with the planetary positions \citep{Zioutas}. A similar correlation also appears in the total electron content of the Earth's atmosphere \citep{Bertolucci}. A tiny amount of extra energy deposition $\sim (0.1-1)$ W/m$^2$ is possible to account for the seasonal variation of the upper stratosphere (i.e. altitude 38.5 km to 47.5 km) \citep{Zioutas,Zhitnitsky}. One recent study has proposed to explain these puzzles by the model of axion quark nuggets (AQNs) dark matter \citep{Zhitnitsky}. The strong interaction between the AQNs and atmospheric particles can provide the required energy deposition. In fact, the interpretation of the stratospheric temperature anomalies is still uncertain and inconclusive. More follow-up analyses are definitely required to argue how the planetary motion is correlated to the reported anomalies. Nevertheless, this issue has initiated a possibility that dark matter might be able to interact with our atmosphere significantly.

In fact, there is another popular dark matter candidate called dark photon dark matter (DPDM) \citep{Redondo,Nelson,Graham,Fabbrichesi}, which can give modest interaction between dark matter and ordinary matter through kinetic mixing with ordinary photons. In this article, I show that the DPDM could be effectively absorbed by our atmosphere due to resonant absorption, provided that only $0.1-1$ W/m$^2$ energy deposition is contributed to the upper stratosphere ($<0.1$\% of the solar energy flux). It can be shown that a very large portion of DPDM could be absorbed by our atmosphere so that only a very small portion of dark matter particles can strike the ground-based direct-detection experiments and axion haloscopes. I will also discuss some interesting and important implications which are somewhat consistent with observations and useful in future dark matter search.

\section{The absorption of dark photon dark matter}
In particle physics, there is a conjectured dark sector parallel to our own \citep{Fabbrichesi}. It contains some unknown states which might constitute the cosmological dark matter in our universe. In view of this, there may exist a kind of photons called dark photons, kinetic mixing with ordinary visible photons. The kinetic mixing between dark photons and ordinary photons provides a possibility to detect dark photons in experiments. The only parameter involved is the mixing parameter $\varepsilon$.

Generally speaking, dark photons can be massless or massive. A very light massive dark photon could be a dark matter candidate (i.e. dark photon dark matter DPDM). DPDM could be produced non-thermally in the early universe as a condensate, like the axion production mechanism \citep{Nelson,Fabbrichesi}. They can also be produced in the inflation era \citep{Graham}. When the Hubble constant drops below the DPDM mass, the DPDM field starts to oscillate and behaves like cold dark matter \citep{Fabbrichesi}. Therefore, in principle, DPDM can form structures and distribute like cold dark matter, which can match the properties observed in large-scale structures and galactic rotation curves. In fact, the DPDM model is currently a popular dark matter model, which has been put on tests in many state-of-the-art experiments and observations \citep{Chiles,An,Hunt,Bajjali,Knirck,Roy,An2}. Some positive evidence has been found to favor the existence of dark photons \citep{Hunt}.

DPDM can interact with ordinary matter through resonant absorption process \citep{Arvanitaki}. In particular, DPDM with energy lower than 1 eV can be absorbed by the vibrational and rotational transition in molecules, such as oxygen and water molecules (like greenhouse effect). The kinetic mixing between DPDM and ordinary photons provides a possibility for the resonant capture of DPDM by an atom or a molecule. For an ordinary photon with frequency $\nu$, the cross section for vibrational and rotational resonant absorption can be written in the following Lorentzian form \citep{Barton,Zak}
\begin{eqnarray}
\sigma_r&=&\frac{Ac^2g \exp(-E/kT)[1-\exp(-h\nu/kT)]}{8\pi \nu^2Q(T)} \nonumber\\
&& \times \left[\frac{\Delta \nu}{\pi} \frac{1}{(\nu-\nu_0)^2+(\Delta \nu)^2} \right],
\end{eqnarray}
where $A$ is the Einstein coefficient, $g$ is the degeneracy factor, $E$ is the initial energy level, $T$ is the temperature of the molecules, $Q(T)$ is the molecular partition function, $\Delta \nu$ is the frequency width of the absorption, and $\nu_0$ is the resonant frequency. The cross section for DPDM can be obtained in terms of the mixing parameter $\varepsilon$ as \citep{Fabbrichesi}:
\begin{equation}
\sigma_A=\varepsilon^2 \sigma_r.
\end{equation}
Assuming there exists a resonant frequency $\nu_0$ in the target molecules which is very close to the DPDM frequency $\nu=m_Ac^2/h$, where $m_A$ is the mass of DPDM. The cross section of DPDM becomes
\begin{equation}
\sigma_A \approx \frac{Ac^2 \varepsilon^2}{8\pi^2 \nu^2 \Delta \nu} \left[1-\exp \left(-\frac{h\nu}{kT} \right) \right] \left[\frac{g \exp(-E/kT)}{Q(T)} \right].
\end{equation}
The Einstein coefficient $A$ is close to the reciprocal of decay life time $(\Delta \tau)^{-1}$. Therefore, based on the uncertainty principle, we have $A \sim (\Delta \tau)^{-1} \sim \Delta \nu$. At the stratosphere, the temperature at altitude $\sim 40$ km is about 250 K. By taking $g \exp(-E/kT) \approx 1$ and $Q(T) \approx 44$ using water molecules as an example \citep{Harris}, we get
\begin{equation}
\sigma_A=4.43 \times 10^{-16} \varepsilon^2 \left(\frac{m_A}{1~{\rm eV}} \right)^{-2} \left[1-\exp \left(-\frac{46.4m_A}{1~\rm eV} \right) \right]~{\rm m^2}.
\end{equation}

As dark matter particles are moving inside the Milky Way galaxy as well as the solar system, we expect that there is a dark matter flux passing through our Earth. The dark matter flux is given by \citep{Neufeld}
\begin{equation}
\Phi=6\times 10^{19}\left(\frac{\rm 1~eV}{m_A} \right) \left(\frac{\rho_{DM}}{\rm 0.3~GeV/cm^3} \right) \left(\frac{\langle v \rangle}{\rm 200~km/s} \right)~{\rm m^{-2}~s^{-1}},
\end{equation}
where $\rho_{DM}$ is the DPDM density at the solar position and $\langle v \rangle$ is the average velocity of DPDM. Assuming DPDM follows a Maxwell-Boltzmann distribution, the average velocity of DPDM is:
\begin{equation}
\langle v \rangle=\int_0^{\infty}\frac{4}{\sqrt{\pi}v_p^3}v^3e^{-v^2/v_p^2}dv=\frac{2v_p}{\sqrt{\pi}},
\end{equation}
where $v_p$ is the characteristic velocity. When DPDM is passing through the Earth's atmosphere, we expect that the absorption process of DPDM by the molecules in the atmosphere would occur. If we assume that all DPDM is absorbed by the atmosphere, the maximum energy flux contributed by DPDM would be
\begin{equation}
\dot{E}=\Phi m_Ac^2=9.6 \left(\frac{\rho_{DM}}{\rm 0.3~GeV/cm^3} \right) \left(\frac{\langle v \rangle}{\rm 200~km/s} \right)~{\rm W/m^2}.
\end{equation}
Taking $\rho_{DM}=0.297$ GeV cm$^{-3}$ constrained from the Gaia data \citep{Labini} and $v_p=270$ km/s \citep{Acevedo}, we get $\dot{E}=14.6$ W/m$^2$. This energy flux is $\sim 10-100$ times of the required energy deposition to explain the temperature anomaly in our stratosphere. 

Although our ultimate goal is not to account for the alleged stratospheric temperature anomalies, we take the required energy deposition $0.1-1$ W/m$^2$ between the stratospheric layers at latitude 38.5-47.5 km reported in \citet{Zioutas,Zhitnitsky} as our reference to analyze the possible absorption. This tiny additional energy deposition is approximately equivalent to producing a maximum temperature variation of 2.5 K in the upper stratosphere \citep{Zioutas}. Considering the optical depth of DPDM and assuming the fraction of the absorption molecules containing the resonant frequency $\nu_0 \approx \nu$ is close to one. The optical depth for each DPDM particle traveling from the vertical height $z=\infty$ to an altitude $z=h$ is
\begin{equation}
\tau(h)=\frac{1}{\bar{m}} \int_{\infty}^h \rho_{\rm atm}(z) \sigma_A dz,
\end{equation}
where $\bar{m}=2.4 \times 10^{-26}$ kg is the average mass of an air molecule in our atmosphere and $\rho_{\rm atm}$ is the mass density of air. By applying the barometric formula, the mass density of air in troposphere and stratosphere is respectively given by
\begin{equation}
\rho_{\rm atm}(z)=\rho_b \left[ \frac{T_b-(z-z_b)L_b}{T_b} \right]^{\frac{g_0M}{RL_b}-1},
\end{equation}
and
\begin{equation}
\rho_{\rm atm}(z)=\rho_b \exp \left[\frac{g_0M(z-z_b)}{RT_b} \right],
\end{equation}
where $\rho_b$, $T_b$ and $L_b$ are the standard density, temperature, and lapse rate respectively in different successive layers, $g_0=9.807$ m/s$^2$ is the surface gravitational acceleration, $R=8.314$ N m mol$^{-1}$ K$^{-1}$ is the universal gas constant and $M=0.02896$ kg/mol is the molar mass of air. In Fig.~1, we plot the variation of $\tau(h)$ for the optical depth constrained by the stratospheric anomalies.

Now we consider the stratospheric absorption between $h=47.5$ km to $h=38.5$ km, the atmospheric layer where the energy deposition flux is calculated. By integrating $\int \rho_{\rm atm}(z)dz$ using the barometric formula, the optical depth from $h=47.5$ km to $h=38.5$ km is 
\begin{eqnarray}
\Delta \tau&&=\tau(38.5)-\tau(47.5) \nonumber\\
&&=5.30 \times 10^{11} \varepsilon^2 \left(\frac{m_A}{1~\rm eV} \right)^{-2} \left[1-\exp \left(-\frac{46.4m_A}{1~\rm eV} \right) \right].
\end{eqnarray}
Since the extra energy deposition in this particular layer is $\dot{Q} \sim 0.1-1$ W/m$^2$ \citep{Zioutas,Zhitnitsky}, we get $\Delta \tau=\dot{Q}/\dot{E}=0.0068-0.068$. This can give a narrow possible parameter space of $\varepsilon$ and $m_A$ for the DPDM model: 
\begin{equation}
\varepsilon^2 \left(\frac{m_A}{\rm 1~eV} \right)^{-2} \left[1-\exp \left(-\frac{46.4m_A}{\rm 1~eV} \right) \right]=(1.28-12.8) \times 10^{-14}.
\end{equation}
We plot the allowed $\varepsilon$ against $m_A$ (in red shaded band) in Fig.~2. We can see that a large region of $\varepsilon-m_A$ parameter space are ruled out by the astrophysical and cosmological bounds, except for the narrow range of dark matter mass $m_A \sim 0.0001-0.001$ eV. In other words, if the stratospheric anomalies are real, DPDM with $m_A \sim 0.0001-0.001$ eV can simultaneously explain the anomalies and satisfy the current astrophysical (e.g. based on JWST observations and solar constraints) \citep{Vinyoles,Li2,An2} and cosmological (e.g. based on cosmic microwave background data) \citep{Arias,Witte} bounds. Here, we also include the excluded parameter space of solar dark photons constrained by the XENON1T experiment in \citet{Aprile4} for comparison because the produced dark photons are relativistic so that the absorption cross section would be different from that for DPDM. Note that since there is a huge amount of discrete resonant frequencies for molecules, the actual limit on $\varepsilon$ should be in the form of many discrete sharp lines. Nevertheless, in Fig.~2, we only consider the case of resonant absorption so the resultant limit of $\varepsilon$ in Fig.~2 appears continuous. In Fig.~3, by considering the rotational excitation energies in oxygen molecules $\rm O_2$ \footnote{The data of the rotational excitation photon frequencies for oxygen molecules O$_2$ are obtained from the Molecular Spectral Databases from National Institute of Standards and Technology.}, we plot the lower limits of $\varepsilon$ for $m_A=0.0001-0.004$ eV. The positions of the sharp line limits are corresponding to the possible resonant energies for rotational excitation due to DPDM absorption. For vibrational absorption, the corresponding $m_A$ for oxygen molecules is greater than 0.24 eV, which is ruled out by other constraints. In general, other air molecules such as nitrogen and water molecules could also contribute to DPDM absorption. Here, the lower limit of $\varepsilon$ shown in Fig.~3 is just a particular example demonstrating the line shaped limit. All the possible parameter space of $\varepsilon-m_A$ is enclosed by the red shaded region in Fig.~2. 

On the other hand, in Fig.~1, we can see that the total optical depth at $h=0$ is $\tau(0) \sim 2.2-22$, which means that almost all DPDM particles would be absorbed in our atmosphere. Therefore, the dark matter flux that can continue to pass through the Earth crust and also the resonant cavity apparatus used in the detection experiments would be largely suppressed. This might explain why we cannot detect any signal of dark matter using the direct-detection experiments and cavity experiments. The existing experimental bounds on $\varepsilon$ \citep{Caputo} would need to be significantly revised because the actual number of dark matter particles passing through the experimental apparatus is much overestimated. Therefore, in this context, the bounds on $\varepsilon$ using direct-detection experiment \citep{Aprile2} and cavity experiments \citep{ADMX2} are no longer viable so we do not include the corresponding bounds in Fig.~2. Nevertheless, the astrophysical and cosmological bounds are still applicable for constraining $\varepsilon$ as the corresponding data are not affected by our atmosphere.   

\section{Other important implications}
There are some interesting implications if DPDM can be effectively absorbed by ordinary matter. In the interstellar medium, the number density is too low ($n \sim 1-10$ cm$^{-3}$) so that the absorption of DPDM is not effective. Nevertheless, in planets, stars or dense clouds, the heating of dark matter might be significant. Also, as a larger DPDM flux can be found in higher altitude (not yet absorbed), there is a new possible way for direct dark matter detection. 

\subsection{Heating of planets}
In planets and stars, the density is high enough for DPDM to be absorbed. A large amount of DPDM would first heat up the atmosphere and then the planetary or stellar body. However, in stars, the energy generated by nuclear fusion is much larger than the dark matter heating. Therefore, the stellar heating due to DPDM is negligible. Nevertheless, in planets, the power absorbed from the host star might be comparable to the dark matter heating rate. Consider our solar system. The solar luminosity flux goes like $1/r^2$ while the dark matter heating flux is almost constant (neglecting the effect of planetary motion). In Fig.~4, we plot the energy flux as a function of the distance from the sun. One can see that dark matter heating becomes important in Jovian planets. Coincidently, many studies have indicated that there is internal heating found in Jupiter (with the rate $\dot{Q}=7.485\pm 0.163$ W/m$^2$) \citep{Li}, Saturn \citep{Ingersoll}, Neptune \citep{Markham}, and Pluto \citep{Witze}. Also, the high temperature at the center of Uranus might indicate a non-adiabtic process \citep{Neuenschwander}. Internal heating can also be found recently in exoplanets \citep{Welbanks}. Although there are several possible known internal heating sources such as tidal heating and radioactive decay, heating the planets by DPDM may also be a possible origin. Moreover, dark matter heating might also contribute to the energy input in the planetary atmosphere to cause temperature inversion. Such a temperature inversion can be seen in Earth's atmosphere and Pluto's atmosphere \citep{Gladstone}. Therefore, investigating the heating of planets and planetary atmosphere is a possible way to test the DPDM proposal.

\subsection{Supermassive dark star}
It has been suggested that some dark star heated by dark matter might exist in the early epoch. The original proposal of the supermassive dark star suggests that a massive gas halo can be ignited by WIMP dark matter annihilation \citep{Spolyar,Freese}. The energy input can give a sufficient temperature $T>10^4$ K and luminosity $L \ge 10^9L_{\odot}$ for us to observe even in redshift $>10$ \citep{Freese}. Recently, a study claims that 3 possible supermassive dark stars (JADES-GS-z13-0, JADES-GS-z12-0, JADES-GS-z11-0) might have been observed \citep{Ilie}. This reveals a possibility of dark matter heating of a massive gas halo in the early universe. In the DPDM context, the heating of DPDM can also provide sufficient energy to give the required luminosity. The adiabatic contraction would give a high ambient dark matter density according to the relation: $\rho_{DM} \sim 5(n_H/{\rm cm^3})^{0.81}$ GeV/cm$^3$ \citep{Spolyar}. For the average baryonic density $n_H \sim 10^{13}$ cm$^{-3}$, we have $\rho_{DM} \sim 10^{11}$ GeV/cm$^3$. The size of a supermassive dark star can be as large as 10 AU \citep{Ilie}. Therefore, the total dark matter mass inside the supermassive dark star is $M_{DM} \sim 10^{-3}M_{\odot}$. Taking the average velocity of the gas particle as $v=\sqrt{3kT/m_H}$ with $T=10^4$ K and assuming all gas particles having the resonant absorption frequency $\nu_0 \approx \nu$, the heating rate of DPDM is
\begin{equation}
\dot{Q}=M_{DM}n_H \sigma_Avc^2>10^9L_{\odot}.
\end{equation}
This shows that the heating of DPDM can also provide the required luminosity for the supermassive dark star observed recently. As the dark matter mass depletion rate is $\dot{M}_{DM}=\dot{Q}/c^2>6.7 \times 10^{-5}M_{\odot}$/yr, the accretion rate of dark matter mass larger than the depletion rate is required to maintain the luminosity for a long time.

\subsection{Direct detection of dark matter at high altitude}
Since we have shown that only a small amount of DPDM particles would be absorbed in the stratosphere and mesosphere, this provides a window for us to detect dark matter at high altitude. For example, in the stratosphere ($h>20$ km), less than 40\% of DPDM would be absorbed by the stratosphere based on our assumed energy deposition rate. Therefore, detecting the signals of dark matter in the stratosphere would be much more effective. One can design a cavity experiment performed in the stratosphere using a balloon. This can equivalently push the limit of the mixing parameter $\varepsilon$ by more than 100 times compared with the experiments performed at the sea-level if $\tau(0)=10$. Based on the current constraints of the haloscope experiments \citep{Caputo}, pushing the limits by 100 times can definitely make the DPDM signal detectable because the revised projected limits would overlap with our constrained $\varepsilon-m_A$ parameter space.

\section{Discussion and Conclusion}
In this article, I have argued that the atmospheric resonant absorption of DPDM is possible. If the so-called stratospheric temperature anomalies are real phenomena, then our proposal could account for the required energy deposition. The resonant absorption can originate from the rotational and vibrational eigenstate transition of diatomic and triatomic molecules in the atmosphere \citep{Arvanitaki}. There are many possible resonant absorption energies $<0.01$ eV in atmospheric molecules, such as oxygen \citep{Vleck,Toureille}. In fact, many recent studies have turned to consider DPDM as a highly probable candidate of dark matter so the DPDM model has been put on different experimental tests \citep{Chiles,An,Hunt,Bajjali,Knirck,Roy,An2}. Some recent experimental results even favor the existence of dark photon \citep{Hunt}. Based on our assumptions, we can get a narrow constrained DPDM mass range $m_A \sim 0.0001-0.001$ eV and the viable range of the mixing parameter $\varepsilon$ without violating current astrophysical and cosmological bounds. Note that the cross section $\sigma_A$ in Eq.~(4) involves a few approximations, including the value of the partition function assumed. Nevertheless, the order of magnitudes of the allowed $\varepsilon$ provides a viable window for strong interaction between dark matter and ordinary matter. It also justifies the possibility of having atmospheric absorption of DPDM. 

Previously, we often believe that the interaction between dark matter particles and ordinary matter is extremely small. This is because we have not detected any interaction signal between dark matter and ordinary matter in the ground-based direct-detection experiments. However, this might be a selection effect in which most of the dark matter particles have already been absorbed by the atmosphere. Here, we have discussed a possibility that most of the dark matter particles are absorbed by the atmosphere based on the DPDM model, without adding any new assumptions to the existing DPDM model. Even if the stratospheric temperature anomalies are false or the anomalies do not originate from DPDM heating, our analysis still shows that effective absorption of DPDM in our atmosphere is possible. It would overturn our usual assumption of extremely weak interaction between dark matter and ordinary matter and revive our hope of detecting dark matter directly. Moreover, we have discussed some interesting implications for the effective absorption of DPDM, such as heating of Jovian planets, igniting supermassive dark star, and performing the cavity experiment at high altitude. These initiate new research directions which can help verify the DPDM model and solve the long-lasting dark matter problem. 

\begin{figure}
\begin{center}
\vskip 5mm
\includegraphics[width=150mm]{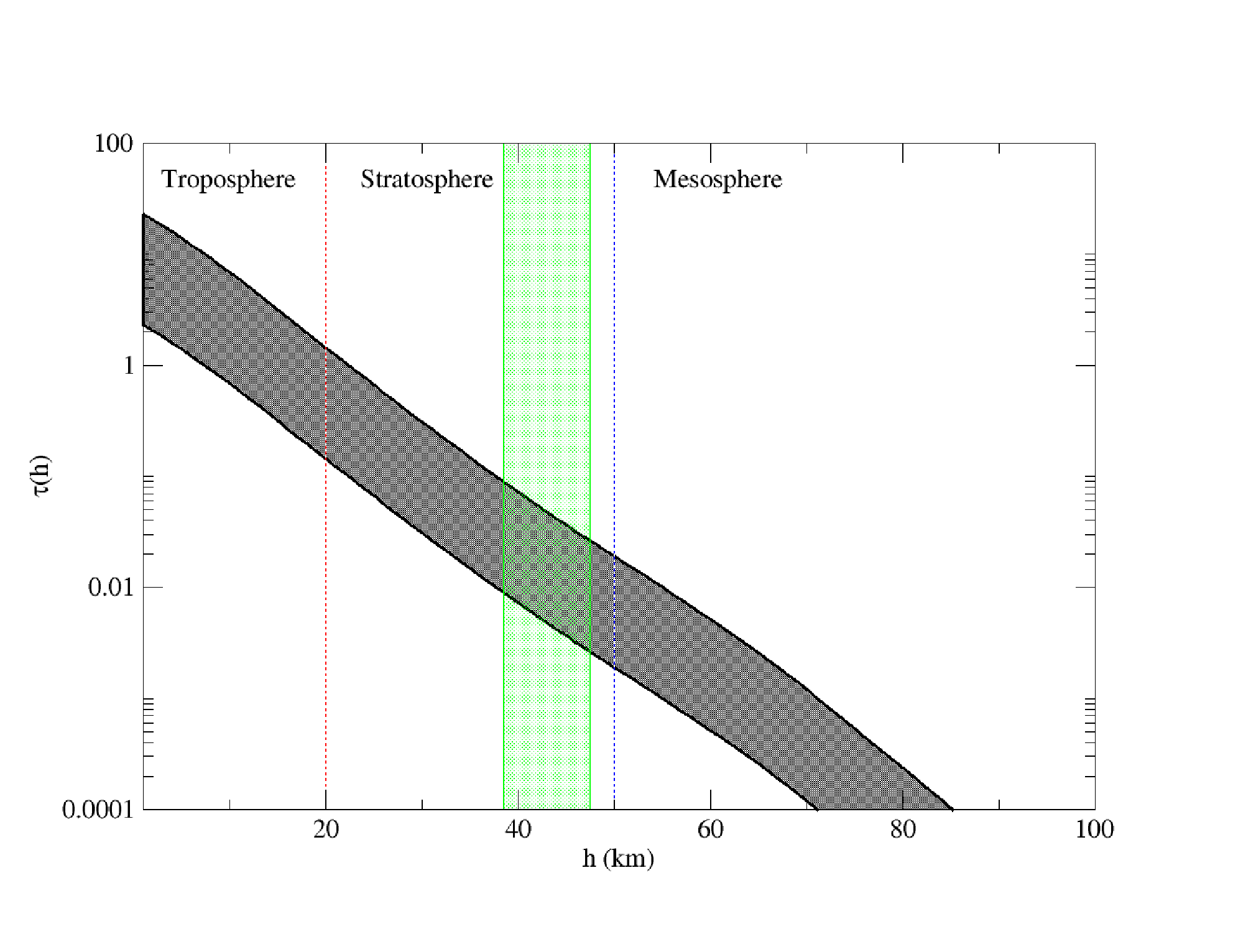}
\caption{The black shaded band represents the possible range of the optical depth of DPDM $\tau(h)$ from $z=\infty$ to $z=h$ (assumed $\dot{Q}=0.1-1$ W/m$^2$). The green shaded area indicates the layer ($h=38.5-47.5$ km) of the stratospheric temperature anomaly reported in \citet{Zioutas}. Here, we have assumed $\varepsilon^2(m_A/1~{\rm eV})^{-2}[1-\exp(-46.4(m_A/1~{\rm eV}))]=(1.28-12.8)\times 10^{-14}$. The dotted lines demarcate different layers of our atmosphere.}
\label{Fig1}
\end{center}
\end{figure}

\begin{figure}
\begin{center}
\vskip 5mm
\includegraphics[width=150mm]{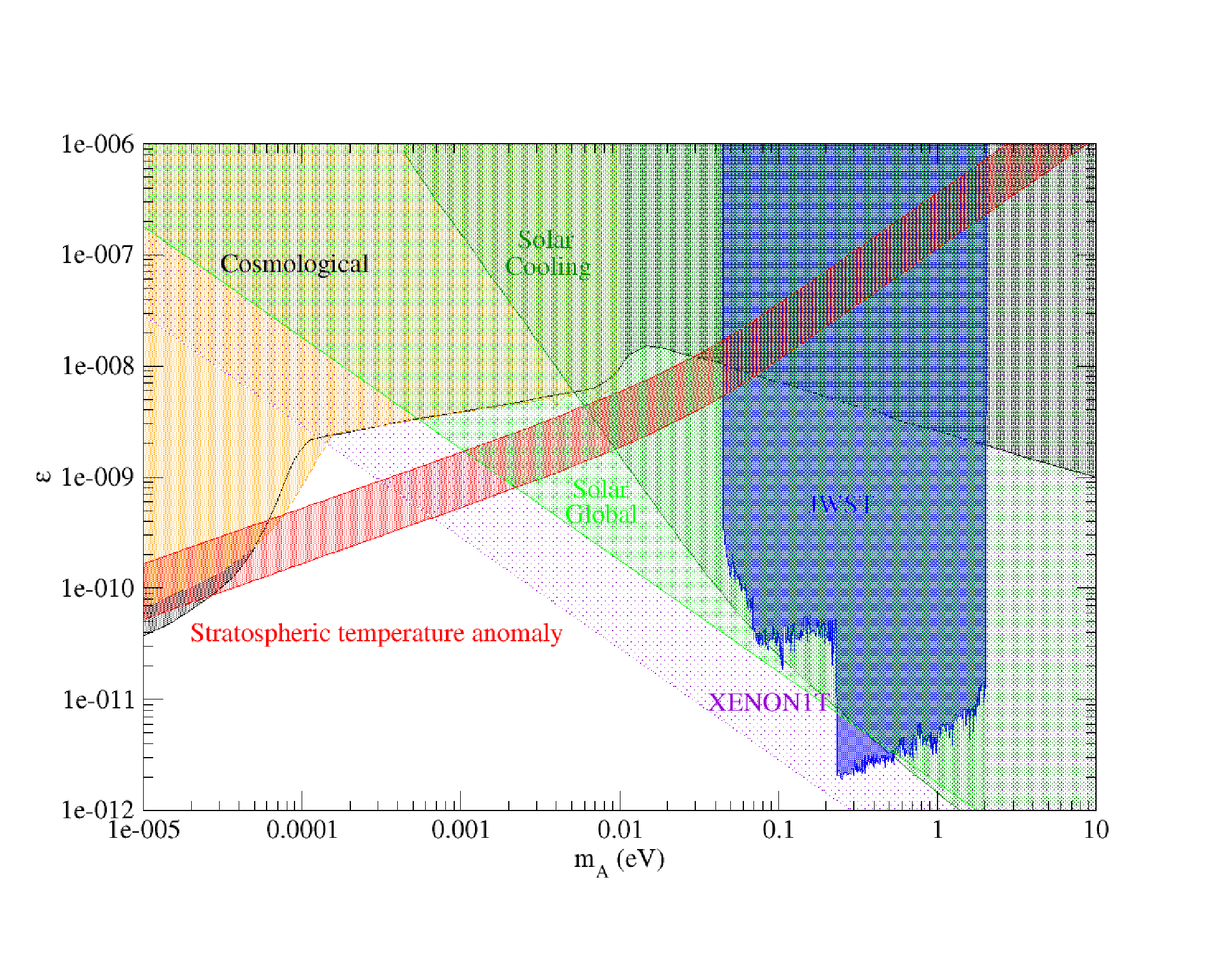}
\caption{The red shaded band represents the allowed parameter space of $\varepsilon$, assuming $\dot{Q}=(0.1-1)$ W/m$^2$. The black and orange shaded regions represent the ruled out parameter space based on the cosmological observations in \citet{Arias} and \citet{Witte} respectively. The blue, green and dark green shaded regions represent the ruled out parameter space based on the JWST observations \citep{An2}, solar global analysis \citep{Vinyoles}, and solar cooling constraints \citep{Li2} respectively. The violet shaded region represents the ruled out parameter space of solar dark photons based on XENON1T experiment \citep{Aprile4}.}
\label{Fig2}
\end{center}
\end{figure}

\begin{figure}
\begin{center}
\vskip 5mm
\includegraphics[width=150mm]{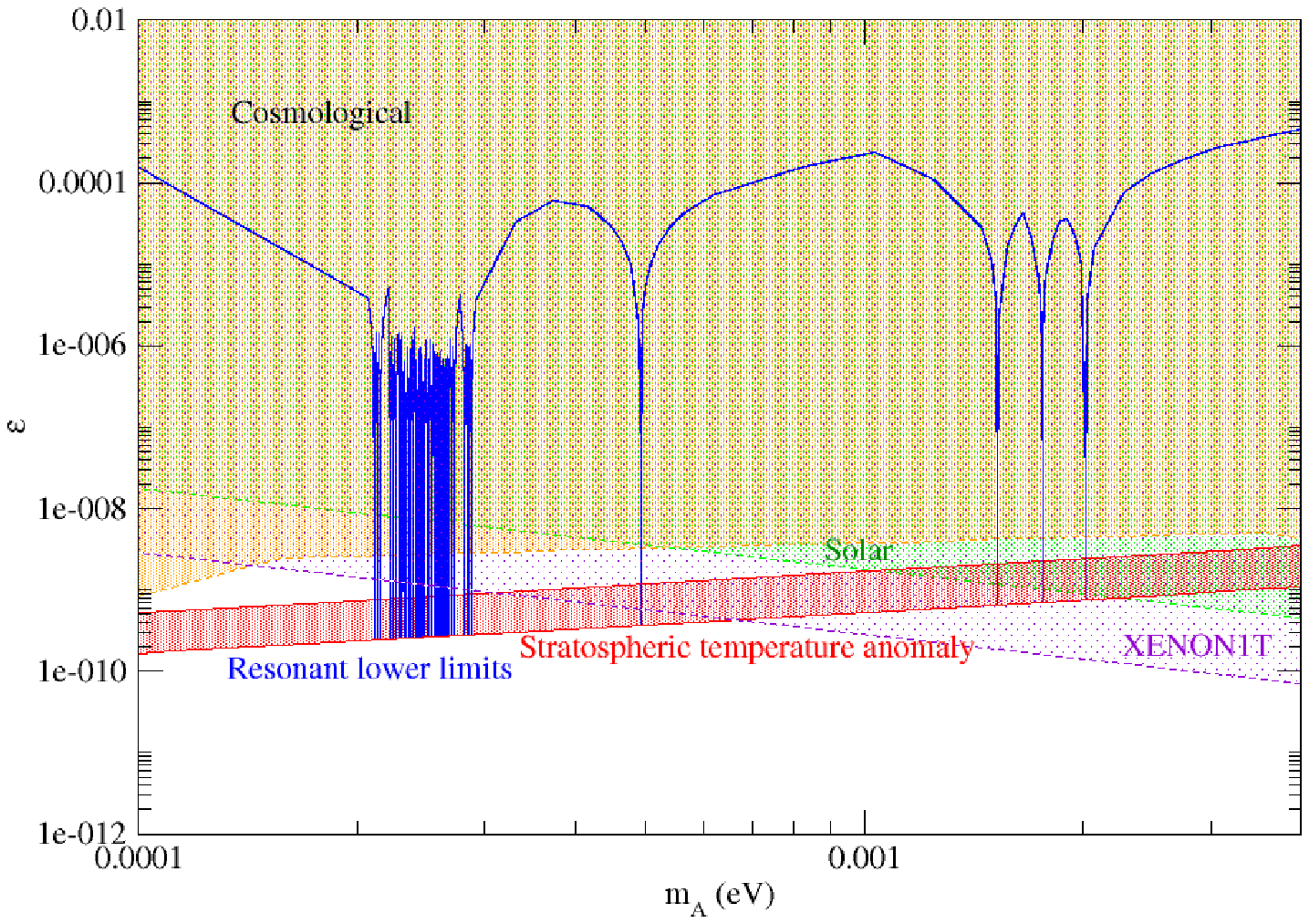}
\caption{The blue solid line indicates the lower limit of $\varepsilon$, assuming the resonant absorption (rotational excitation) with $\dot{Q}=0.1$ W/m$^2$ originated from oxygen molecules. The red shaded band represents the continuous resonant constraint of $\varepsilon$ shown in Fig.~2. The orange, green and violet shaded regions represent the ruled out parameter space based on the cosmological observations in \citet{Witte}, solar global analysis in \citet{Vinyoles}, and solar dark photon constraints in XENON1T experiment \citep{Aprile4}.} 
\label{Fig3}
\end{center}
\end{figure}

\begin{figure}
\begin{center}
\vskip 5mm
\includegraphics[width=140mm]{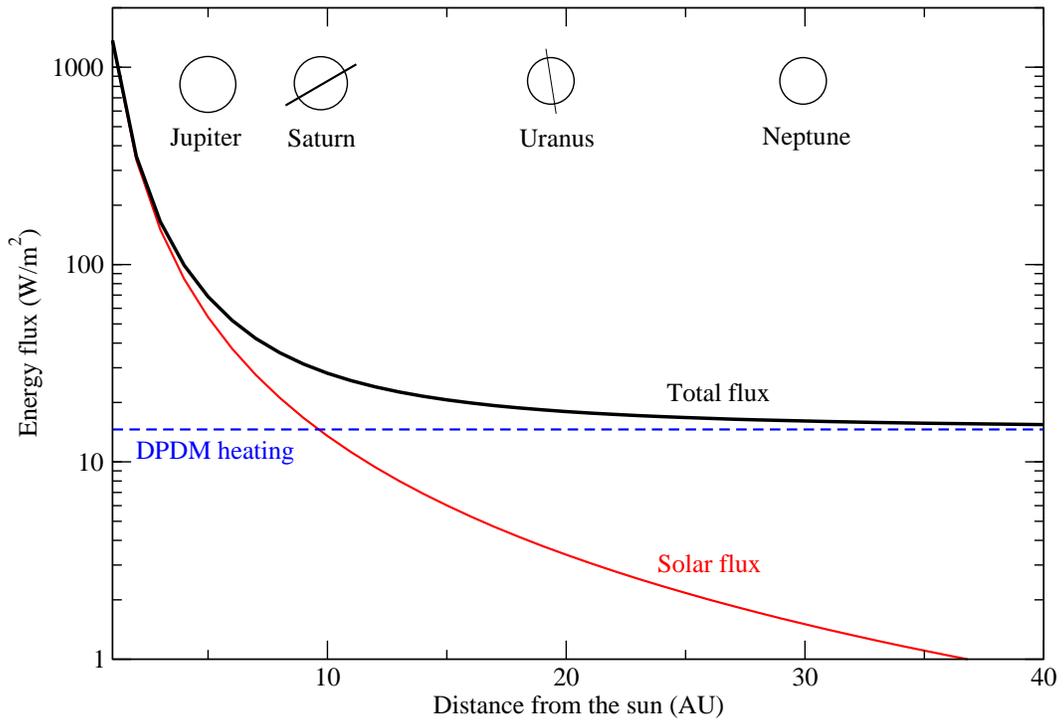}
\caption{The red solid line and blue dashed line respectively represent the solar flux and the heating flux of DPDM. The black solid line represents the total flux by adding the solar flux with the DPDM heating flux. Here, we have neglected the effect of bond albedo for the planets. The positions of the Jovian planets are shown in the figure.}
\label{Fig4}
\end{center}
\end{figure}

\section{Acknowledgements}
I thank the anonymous referee for the useful comments. The work described in this paper was partially supported by the Dean's Research Fund of The Education University of Hong Kong (0400W) and the grants from the Research Grants Council of the Hong Kong Special Administrative Region, China (Project No. EdUHK 18300922 and EdUHK 18300324).

\end{document}